\documentclass[12pt]{iopart}

\usepackage{hyperref}
\usepackage[english]{babel}
\usepackage[T1]{fontenc}
\usepackage[latin1]{inputenc}
\usepackage[dvips]{graphicx}
\usepackage{iopams}
\usepackage{setstack}
\usepackage{mathenv}
\usepackage{array}
\usepackage{bm}
\usepackage{fancyhdr}
\usepackage{makeidx}
\usepackage{subfigure}
\usepackage{babel}
\usepackage[v2]{xy}
\usepackage{epsfig}
\usepackage{amsfonts}
\usepackage{amssymb}
\usepackage{color}
\hypersetup{backref=true,pagebackref=true,hyperindex=true,colorlinks=true,breaklinks=true,urlcolor=
blue,linkcolor=blue,bookmarks=true,bookmarksopen=true}

\begin{document}

\title{Description of current-driven torques in magnetic tunnel junctions}
\date{\today}
\date{\today}
\author{A. Manchon$^1$, N. Ryzhanova$^{1,2}$, A. Vedyayev$^{1,2}$, M. Chschiev$^3$ and B. Dieny$^1$}
\address{$^1$ SPINTEC, URA 2512 CEA/CNRS, CEA/Grenoble, 38024 Grenoble Cedex 9, France}
\address{$^2$ Department of Physics, Lomonosov University, Moscow, Russia}
\address{$^3$ MINT Center, University of Alabama, P.O. Box 870209, Tuscaloosa, Alabama, USA}
\ead{aurelien.manchon@m4x.org}

\begin{abstract}
A free electron description of spin-dependent tranport in magnetic tunnel junctions with non collinear magnetizations is presented. We investigate the origin of transverse spin density in tunnelling transport and the quantum interferences which give rise to oscillatory torques on the local magnetization. Spin transfer torque is also analyzed and an important bias asymmetry is found as well as a damped oscillatory behaviour. Furthermore, we investigate the influence of the {\it s-d} exchange coupling on torque in particular in the case of half-metallic MTJ in which the spin transfer torque is due to interfacial spin-dependent reflections.
\end{abstract}
\submitto{\JPCM}
\maketitle
\section{Introduction}

The theoretical demonstration of spin transfer torque in metallic spin valves (SV) ten years ago \cite{slonc} gave a new breath to giant magnetoresistance related studies \cite{GMRreview}, promising exciting new applications in non-volatile memories technology \cite{applSTT} and radio-frequency oscillators \cite{rfSTT}. A number of fundamental studies in metallic spin valves revealed the different properties of spin torque and led to a deep understanding of current-induced magnetization dynamics \cite{katine,kiselev,rippard,urazhdin,alhajdarwish}. Particularly, several theoretical studies described the structure of the torque in metallic magnetic multilayers and showed the important role of averaging due to quantum interference, spin diffusion and spin accumulation \cite{stiles02,autres,jpcm}.\par

Since the first experimental evidence of spin-dependent tunnelling \cite{jullieremeservey}, magnetic tunnel junctions (MTJs) have attracted much attention because of the possibility to obtain large tunnelling magnetoresistance (TMR) at room temperature \cite{moodera}. The possibility to use MTJs as sensing elements in magnetoresistive heads, as non-volatile memory elements or in reprogrammable logic gates has also stimulated a lot of technological developments aiming at the optimization of MTJs' transport properties and their implementation in silicon-based circuitry \cite{applSTT,ieee}. Because of these applications, MTJs have been intensively studied and the role of interfaces \cite{leclair}, barrier \cite{sharma}, disorder \cite{tsymbal98} and impurities \cite{imp} have been addressed in many publications \cite{reviewTMR}. The recent achievement of current-induced magnetic excitations and reversal in MTJs \cite{huaifuchs} has renewed the already very important interest of the scientific community in MTJs.\par

The observation of spin transfer torque in low RA (resistance area product) MTJ using amorphous \cite{huaifuchs} or cristalline barriers \cite{ieee,ideka} opened new questions about the transport mechanism in MTJ with non collinear magnetizations orientations. As a matter of fact, whereas the current-perpendicular-to-plane (CPP) transport in SV is mostly diffusive and governed by spin accumulation and relaxation phenomena \cite{autres,jpcm}, spin transport in magnetic tunnel junctions is mainly ballistic and governed by the coupling between spin-dependent interfacial densities of states: all the potential drop occurs within the tunnel barrier. J. C. Slonczewski first proposed a free electron model of spin transport in a MTJ with an amorphous barrier \cite{slonc89}, deriving TMR, spin transfer torque (STT) and zero bias interlayer exchange coupling (IEC). This first model only considered electrons at Fermi energy, neglecting all non-linear tunnel behaviour (consequently, current-induced IEC was found to be zero). More recently, the author proposed a more general model based on Bardeen's Transfer Matrix (BTM) method \cite{slonc05}. Another group presented at tight-binding model (TB) of a MTJ, giving more realistic band structures than the usual free electron model \cite{theo,kalitsov}. These studies showed that spin torque should present an important bias asymmetry and the dissipative part of IEC (also called current-induced effective field) should be of the same order of magnitude than STT with a quadratic dependence on the bias voltage \cite{theo}. Finally, we note that in the same spirit as Ref. \cite{magnons}, Levy and Fert studied the role of hot electrons-induced magnons on STT in MTJ \cite{levyfert}. In recent experiments, the important relative amplitude of current-induced effective field compared to the spin torque term has been verified \cite{kubota,Petit,Sankey} but the role of magnons is still under investigation (in the first experiment the current-induced magnetization reversal occured while the TMR was quenched by magnons emissions \cite{huaifuchs}). These specific features show that tunnelling transport has a strong influence on spin transfer torque characteristics.\par

We recently presented \cite{jpcm} a description of spin-dependent transport in a MTJ treated in a free-electron assumption, based on Keldysh non-equilibrium technique \cite{keldysh} applied to a MTJs with an amorphous barrier (such as AlO$_x$). This method is close to Ref. \cite{slonc89}, although more general since we consider the contribution of all electrons lying below the Fermi energy. In the present article, we focus on the anatomy of spin transfer torque in such a MTJ, paying attention to the origin of the specific characteristics of this torque in the particular case of MTJ. The paper is organized as follows. In section \ref{s:torques}, we give a reminder of the origin of spin transfer torque, and the way to calculate it. In section \ref{s:keldysh}, the formulation of spin-dependent currents and torques in non-equilibrium Green function formalism (Keldysh formalism) is developed. Section \ref{s:results} presents the results of the model and describe the anatomy of spin torque in a magnetic tunnel junction, underlying the role of tunnelling process.\par

\section{\label{s:torques} Current-induced torques}

All along this paper, we consider the {\it s-d} model in which {\it s}-electrons are itinerant and {\it d}-electrons are localized and give rise to the local magnetization of the ferromagnet. We also assume that the {\it d} local moments remain stationnary. This model applies to the electron structures of ferromagnetic electrodes whose compositions lie on the negative slope side of the Slater-Néel-Pauling curve \cite{stearns73} (Ni, Co, NiFe, CoFe). No spin flip is taken into account. In a magnetic tunnel junction composed of two semi-infinite ferromagnets separated by a tunnel barrier (see Fig. 1), majority spins and minority spins refer to the electron spin projection in the left ferromagnet respectively parallel or antiparallel to the local magnetization. In this framework, the motion of {\it s}-like electrons in a ferromagnet is represented by the non-relativistic Hamiltonian including {\it s-d} coupling:

\begin{equation}\label{e:sdh}
H=\frac{p^2}{2m}+U({\bf r})+J_{sd}\left(\overrightarrow{\sigma}.\overrightarrow{S}_d\right)
\end{equation}
where the first and second terms are the kinetic and potential energies, the third term is the {\it s-d} exchange energy, $\overrightarrow{S}_d$ being a unit vector collinear to the local magnetization due to the localized {\it d}-electrons, $J_{sd}$ the {\it s-d} exchange constant and $\overrightarrow{\sigma}$ is the vector of Pauli matrices in spin space. After some algebra \cite{jpcm,sun}, it is possible to derive the equation of continuity of the spin density :

\begin{equation}\label{e:spincont}	\frac{d}{dt}\overrightarrow{s}\left(\bm{r},t\right)=\frac{\hbar}{2}\{\frac{d}{dt}\Psi^*\overrightarrow{\sigma}\Psi+\Psi^*\overrightarrow{\sigma}\frac{d}{dt}\Psi\}
\end{equation}
where $\Psi=\left(\Psi^\uparrow,\Psi^\downarrow\right)$ is an arbitrary 2-dimension Hartree-Fock wavefunction. The two dimensions refer to majority $\left(\uparrow\right)$and minority $\left(\downarrow\right)$ spin projections of the Hartree-Fock wavefunction. Here, $\overrightarrow{s}\left(\bm{r},t\right)$ refers to the local spin density (namely the local out-of-equilibrium magnetization due to the itinerant polarized electrons):
\begin{equation}
\overrightarrow{s}\left(\bm{r},t\right)=\Psi^*\left(\bm{r},t\right)\frac{\hbar}{2}\overrightarrow{\sigma}\Psi\left(\bm{r},t\right)\\
\end{equation}

Defining
\begin{equation}
\bm{J}^s=-\frac{\hbar^2}{2m}\ \Im\{\Psi^*\left(\bm{r},t\right)\overrightarrow{\sigma}\otimes\nabla_{\bm{r}}\Psi\left(\bm{r},t\right)\}
\end{equation}
where $\bm{J}^s$ refers to the spin-current density, we obtain, in steady states:
\begin{equation}\label{e:spincont4}	
\nabla_{\bm{r}} \bm{J}^s\left(\bm{r},t\right)=\frac{2J_{sd}}{\hbar}\overrightarrow{S_{d}}\times\overrightarrow{s}\left(\bm{r},t\right)
\end{equation}

Eq. \ref{e:spincont4} implies that the spatial transfer of spin momentum from the itinerant {\it s}-electrons to the localized {\it d}-electrons (left-hand side of Eq. \ref{e:spincont4}) is equivalent to a torque exerted by the transverse spin density on the local magnetization (right-hand side of Eq. \ref{e:spincont4}). This equivalence has been demonstrated by Kalitsov et al. \cite{kalitsov} in magnetic tunnel junctions using Keldysh formalism and TB description.\par

\begin{figure}
	\centering
		\includegraphics{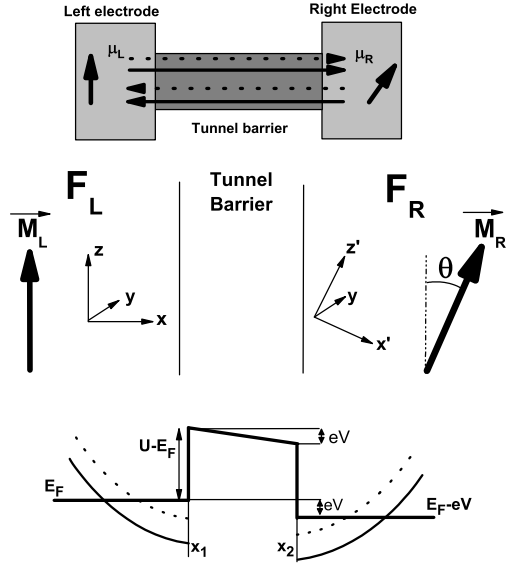}
	\label{fig:figbase}\caption{Schematics of the magnetic tunnel junction with non collinear magnetizations orientation. Top panel: spin-dependent out-of-equilibrium transport in a conductor linking two reservoirs $F_L$ and $F_R$ (whose electrochemical potentials are respectively $\mu_L$ and $\mu_R$) with non collinear magnetizations orientations. The solid arrows represent the majority spins and the dotted arrows represent the minority spins. Middle panel: MTJ with non collinear magnetization orientations. Bottom panel: Corresponding energy profile of the MTJ. In free-electron approximation, the local density of states are parabolic for majority (solid line) and minority (dotted line) electrons with a splitting between the two spin subbands equals to the exchange interaction $J_{sd}$.}
\end{figure}

In the following, we calculate spin transfer torque from the torque exerted by the transverse spin density on the local magnetization. The spirit of our calculation is depicted in the top panel of Fig. 1. The out-of-equilibrium magnetic tunnel junction is modelled by a "conductor" (in the sense that the tunnel barrier is not infinite) linking two magnetic reservoirs ($F_L$ and $F_R$) with non collinear magnetizations and with different chemical potentials $\mu_L$ and $\mu_R$ \cite{datta} ($\mu_L>\mu_R$). A bias voltage $V=(\mu_L-\mu_R)/e$ is applied across this "conductor". One should consider all electrons with majority spins (solid arrows) and minority spins (dotted arrows), originated from left (rightward arrows) and right electrodes (leftward arrows). In low bias limit ($\mu_L\approx\mu_R$), the charge transport can be approximately determined by the electrons originated only from the left electrode at the Fermi energy.\par
In our case (middle panel of Fig. 1), the magnetic tunnel junction is composed of two ferromagnetic layers, $F_L$ and $F_R$ (made of the same material, for simplicity), respectively connected to the left and right reservoirs and separated by an amorphous tunnel barrier. The {\it x}-axis is perpendicular to the plane of the layers and the magnetization of $F_L$ is oriented following z: $\overrightarrow{M}_L=M_L\overrightarrow{z}$. The magnetization $\overrightarrow{M}_R$ of $F_R$ is in the ({\it x,z}) plane and tilted from $\overrightarrow{M}_L$ by an angle $\theta$. In this configuration, the spin density in a ferromagnetic layer possesses three components : $\overrightarrow{m}=(m_x,m_y,m_z)$. In $F_L$ (we obtain the same results considering $F_R$), the transverse components are $m_x=<\sigma^x>$ and $m_y=<\sigma^y>$, where $\sigma^i$ are the Pauli spin matrices and <> denotes averaging over orbital states and spin states, i.e. averaging over electrons energy $E$, transverse momentum $\overrightarrow{\kappa}$ and spin states. Thus, the transverse spin density exerts a torque $\overrightarrow{T}$ on the background magnetization $\overrightarrow{M}_L$ following two axes:

\begin{equation}
\overrightarrow{T}=\frac{J_{sd}}{\mu_B}\overrightarrow{M}_L\times\overrightarrow{m}=\frac{ J_{sd}}{\mu_B}\left[m_x\overrightarrow{M}_L\times\overrightarrow{M}_R-m_y\overrightarrow{M}_L\times\left(\overrightarrow{M}_L\times\overrightarrow{M}_R\right)\right]\label{eq:LLG1}
\end{equation}
One should introduce the previous formula in the usual Landau-Lifshitz-Gilbert equation to describe the modified dynamics of the magnetization $\overrightarrow{M}_L$:
\begin{equation}\label{eq:LLG}
\frac{d\overrightarrow{M}_L}{dt}=\alpha\overrightarrow{M}_L\times\frac{d\overrightarrow{M}_L}{dt}-\gamma\left(\overrightarrow{M}_L\times\overrightarrow{H_{eff}}+\overrightarrow{T}\right)
\end{equation}
where $\alpha$ is the Gilbert damping, $\gamma$ is the gyromagnetic ratio and $\mu_B$ is the Bohr magnetron. The two terms in the right hand side of Eq. \ref{eq:LLG1} stand for two types of torques: $-m_y\overrightarrow{M}_L\times\left(\overrightarrow{M}_L\times\overrightarrow{M}_R\right)$ is the usual STT term (also called in-plane or parallel torque\cite{theo}) whereas $m_x\overrightarrow{M}_L\times\overrightarrow{M}_R$ is the current-driven interlayer exchange coupling (also called field-like torque, out-of-plane or perpendicular torque\cite{theo}). The former vanishes at zero bias, whereas the latter exists even without current \cite{slonc89,theo,kalitsov}. An explanation of the physical nature and origin of these two terms will be given in section \ref{s:results}. The transverse spin density in the left layer is then given by $<\sigma^+>=<\sigma^x+i\sigma^y>$ :
\begin{equation}\label{e:1}
\fl m_x+im_y=<\sigma^+>=<\left(\begin{tabular}{cc}
$\Psi^{*\uparrow}$ & $\Psi^{*\downarrow}$
\end{tabular}\right)\left(
\begin{tabular}{cc}
$0$&$2$ \\
$0$&$0$
\end{tabular}\right)\left(
\begin{tabular}{c}
$\Psi^{\uparrow}$ \\
$\Psi^{\downarrow}$
\end{tabular}\right)>=2<\Psi^{*\uparrow}\Psi^{\downarrow}>
\end{equation}
In other words, STT is given by the imaginary part of $<\sigma^+>$, while IEC is given by its real part. One can understand the product $<\Psi^{*\uparrow}\Psi^{\downarrow}>$ as a correlation function between the two projections of the spin of the impinging electron. In ballistic regime, an electron impinging on a ferromagnet with a spin polarization tilted from the background magnetization will precess around this magnetization \cite{stiles02,kalitsov}. Locally, its two projections $\uparrow$ and $\downarrow$ following the quantization axis (defined by the background magnetization) will be non-zero. Then, the electron will contribute locally to the transverse spin density $m_x$ and $m_y$. If the electron spin is fully polarized parallel or antiparallel to this magnetization, no precession will occur and its contribution to the transverse spin density will be zero.\par

We remind that we defined majority (minority) states as the spin projection parallel (antiparallel) to the magnetization of the left electrode. Then, $<\Psi^{*\uparrow}\Psi^{\downarrow}>$ will be the fraction of electrons whose spin is following $x$ (real part) and $y$ (imaginary part) in spin space.

\section{Formulation of currents and torques\label{s:keldysh}}
\subsection{Keldysh Green functions}

As explained previously, in Keldysh out-of-equilibrium formalism \cite{keldysh,datta}, any physical quantity should be calculated considering the contribution of the electrons originated from the left reservoir {\it and} from the right reservoir (top panel of Fig. 1). Then, an out-of-equilibrium Green function $G(\bm{r},t,\bm{r}',t')$ (or Keldysh Green function) is defined as a superposition of these two contributions:

\begin{equation}\label{e:gmp0}
G\left(\bm{r},t,\bm{r}',t'\right)=f_L\Psi_L\left(\bm{r},t\right)\Psi_L^{*}\left(\bm{r}',t'\right)+f_R\Psi_R\left(\bm{r},t\right)\Psi_R^{*}\left(\bm{r}',t'\right)
\end{equation}
where $\Psi_{L(R)}\left(\bm{r},t\right)$ are the electron wavefunctions originated from the left (right) reservoir at the location $\bm{r}$ and time $t$ and $f_{L(R)}$ are the Fermi distribution functions in the left and right reservoirs.\par

Thus, the Schrödinger equation of the magnetic tunnel junction is:

\begin{equation}\label{e:2}
	H\Psi=\left(\frac{p^2}{2m}+U+J_{sd}\left(\overrightarrow{\sigma}.\overrightarrow{S_d}\right)\right)\left(
\begin{tabular}{c}
$\Psi^{\uparrow}$ \\
$\Psi^{\downarrow}$
\end{tabular}\right)=E\left(\begin{tabular}{c}
$\Psi^{\uparrow}$ \\
$\Psi^{\downarrow}$
\end{tabular}\right)
\end{equation}
where $\overrightarrow{\sigma}$ the vector in Pauli matrices space : $\overrightarrow{\sigma}=(\sigma^x, \sigma^y, \sigma^z)^T$, $E$ is the electron energy, $U$ is the spin-independent potential along the junction:

$$J_{sd}\left(\overrightarrow{\sigma}.\overrightarrow{S_d}\right)=J_{sd}\sigma^z\ \ \ \mbox{and}\ \ \ U=E_F\ \ \ \mbox{for}\ \ \ x<x_1$$
$$J_{sd}\left(\overrightarrow{\sigma}.\overrightarrow{S_d}\right)=0\ \ \ \mbox{and}\ \ \ U(x)=U_0-\frac{x-x_1}{x_2-x_1}eV\ \ \ \mbox{for}\ \ \ x_1<x<x_2$$
$$J_{sd}\left(\overrightarrow{\sigma}.\overrightarrow{S_d}\right)=J_{sd}\left(\sigma^z\cos\theta+\sigma^x\sin\theta\right)\ \ \ \mbox{and}\ \ \ U=E_F-eV\ \ \ \mbox{for}\ \ \ x>x_2$$

We consider that the potential drop occurs essentially within the barrier and we apply a low bias voltage compared to the barrier height ($V<<U/e$). This allows to use WKB approximation to determine the wavefunctions inside the barrier. Furthermore, the free electron approximation implies parabolic dispersion laws which also restricts our study to low bias voltage.\par
To describe the spin-dependent transport within the MTJ, we define the wavefunctions $\Psi^{\sigma'(\sigma)}_i(\bm{r},\epsilon)$, where $\epsilon=E_F-E$ and $E$ is the tunnelling electron energy. $|\Psi^{\sigma'(\sigma)}_i(\bm{r},\epsilon)|^2$ is the probability that an electron originated from electrode $i$, at the energy $\epsilon$, initially in spin state $\sigma$, possesses a spin projection $\sigma'$ at the location $\bm{r}$. For example, an electron initially in majority state, originated from $F_L$, is described by six wavefunctions along the structure:

$$\Psi^{\uparrow(\uparrow)}_L=\frac{1}{\sqrt{k_1}}e^{ik_1x}+b_Le^{-ik_1x}$$
$$\Psi^{\downarrow(\uparrow)}_L=d_Le^{-ik_2x}$$

in the left electrode $F_L$ ($x<x_1$),

$$\Psi^{\uparrow(\uparrow)}_L=\frac{a'_LE(x_1,x)+b'_LE(x,x_1)}{q(x)}$$
$$\Psi^{\downarrow(\uparrow)}_L=\frac{c'_LE(x_1,x)+d'_LE(x,x_1)}{q(x)}$$
where $E(x_i,x_j)=exp\left(\int_{x_i}^{x_j}q(x)dx\right)$, in the tunnel barrier ($x_1<x<x_2$),

$$\Psi^{\uparrow(\uparrow)}_L=a''_Le^{ik_3x}+b''_Le^{ik_4x}$$
$$\Psi^{\downarrow(\uparrow)}_L=c''_Le^{ik_3x}+d''_Le^{ik_4x}$$

in the right electrode $F_R$ ($x>x_2$). $k_{1(2)}$ and $k_{3(4)}$ are the wavevectors for majority (minority) spin projection in the left and right electrodes, whereas $q(x)$ is the spin-independent wavevector inside the tunnel barrier. Connecting the wavefunctions and their derivatives at the interfaces, we obtain the 24 wavefunctions (two spin projections and two reservoirs). These wavefunctions are given in Appendix.\par
In the 2-dimensionnal Hartree-Fock representation, spin-dependent current and spin density are defined using the out-of-equilibrium lesser Keldysh Green function: 

\begin{eqnarray}\label{e:gmp1}
G_{\sigma\sigma'}^{-+}\left(\bm{r},\bm{r}'\right)=&&\int d\epsilon \left(f_L\left[\Psi_L^{\sigma'\left(\uparrow\right)*}\left(\bm{r}'\right)\Psi_L^{\sigma\left(\uparrow\right)}\left(\bm{r}\right)+\Psi_L^{\sigma'\left(\downarrow\right)*}\left(\bm{r}'\right)\Psi_L^{\sigma\left(\downarrow\right)}\left(\bm{r}\right)\right]\right.\nonumber\\
&&\left.+f_R\left[\Psi_R^{\sigma'\left(\uparrow\right)*}\left(\bm{r}'\right)\Psi_R^{\sigma\left(\uparrow\right)}\left(\bm{r}\right)+\Psi_R^{\sigma'\left(\downarrow\right)*}\left(\bm{r}'\right)\Psi_R^{\sigma\left(\downarrow\right)}\left(\bm{r}\right)\right]\right)
\end{eqnarray}
where $f_L=f^0(\epsilon)$ and $f_R=f^0(\epsilon+eV)$, and $f^0(\epsilon)$ is the Fermi distribution at 0 K. For conveniency, we use the mixed-coordinate system $(x,\overrightarrow{\kappa})$, where $\overrightarrow{\kappa}$ is the momentum parallel to the plane and $x$ is the coordinate perpendicular to the plane. With $\bm{r}=(x,\overrightarrow{\rho})$, we get:
\begin{equation}\label{e:gmp2}
G_{\sigma\sigma'}^{-+}\left(\bm{r},\bm{r}'\right)=\frac{a_0}{2\sqrt{\pi}}\int_0^{2\sqrt{\pi}/a_0}e^{i\overrightarrow{\kappa}\left(\overrightarrow{\rho}-\overrightarrow{\rho'}\right)}G_{\sigma\sigma'}^{-+}(x,x')d\overrightarrow{\kappa}
\end{equation}
where $a_0$ is the lattice parameter of the electrodes \cite{cutoff}. Spin transfer torque (STT, $T_{||}$) and interlayer exchange coupling (IEC, $T_{\bot}$) can now be determined from Eq. \ref{e:1}, whereas spin-dependent electrical current densities are calculated from the usual local definition:
\begin{eqnarray}\label{e:defiecstt}
&&T_{\bot}+iT_{||}=\frac{J_{sd}}{\mu_B}<\sigma^+>=2\frac{J_{sd}}{\mu_B}\frac{a_0^3}{(2\pi)^2}\int\int G^{-+}_{\uparrow\downarrow}(x,x,\epsilon)\kappa d\kappa d\epsilon\\\label{eq:mz}
&&m_z=\frac{J_{sd}}{\mu_B}\frac{a_0^3}{(2\pi)^2}\int\int \left[G^{-+}_{\uparrow\uparrow}(x,x,\epsilon)-G^{-+}_{\downarrow\downarrow}(x,x,\epsilon)\right]\kappa d\kappa d\epsilon\\\label{e:defcurrent}
&&J_{\uparrow\left(\downarrow\right)}=\frac{\hbar e}{4\pi m_e}\int\int\left[\frac{\partial}{\partial x}-\frac{\partial}{\partial x'}\right]G^{-+}_{\uparrow\uparrow(\downarrow\downarrow)}(x,x',\epsilon)|_{x=x'}\kappa d\kappa d\epsilon\\
&&J=J_\uparrow+J_\downarrow
\end{eqnarray}
$G^{-+}_{\uparrow\uparrow}(x,x,\epsilon)$ and $G^{-+}_{\downarrow\downarrow}(x,x,\epsilon)$ are the energy-resolved local density-of-states (LDOS) for up- and down-spins respectively, whereas $\int G^{-+}_{\uparrow\uparrow}(x,x,\epsilon)d\epsilon$ and $\int G^{-+}_{\downarrow\downarrow}(x,x,\epsilon)d\epsilon$ give the number of up- and down-electrons at the location $x$ along the structure.\par

\subsection{Calculation of spin transfer torque}

As demonstrated in Eq. \ref{e:spincont4}, it is possible to calculate spin transfer torque from the divergency of spin current density or from the spin density itself. We now demonstrate that this relation holds in our model. Spin current densities and spin density are defined as \cite{stiles02}:

\begin{eqnarray}
m_x=\left[\Psi^{\downarrow}\Psi^{*\uparrow}+\Psi^{\uparrow}\Psi^{*\downarrow}\right]\\
m_y=-i\left[\Psi^{\downarrow}\Psi^{*\uparrow}-\Psi^{\uparrow}\Psi^{*\downarrow}\right]\\
J^s_{x}=-\frac{\hbar^2}{2m}\ \Im\{\Psi^{*\uparrow}\frac{\partial\Psi^{\downarrow}}{\partial x}+\Psi^{*\downarrow}\frac{\partial\Psi^{\uparrow}}{\partial x}\}\\
J^s_{y}=-\frac{\hbar^2}{2m}\ \Re\{\Psi^{*\downarrow}\frac{\partial\Psi^{\uparrow}}{\partial x}-\Psi^{*\uparrow}\frac{\partial\Psi^{\downarrow}}{\partial x}\}
\end{eqnarray}

We evaluate these quantities for electrons originating from the left reservoir in the left electrode ($x<x_1$). The equations are given in Appendix. The spin densities for majority ($\uparrow$) and minority ($\downarrow$) electrons are:

\begin{eqnarray}\label{eq:mxmy}
\fl m_x^\uparrow=8q_1q_2(k_3-k_4)\sin\theta\left(\frac{e^{-i(k_1+k_2)(x-x_1)}-r_1^{*\uparrow}e^{i(k_1-k_2)(x-x_1)}}{den}+c.c.\right)\\
\fl m_x^\downarrow=8q_1q_2(k_3-k_4)\sin\theta\left(\frac{e^{-i(k_1+k_2)(x-x_1)}-r_1^{*\downarrow}e^{-i(k_1-k_2)(x-x_1)}}{den}+c.c.\right)\\
\fl m_y^\uparrow=-8iq_1q_2(k_3-k_4)\sin\theta\left(\frac{e^{-i(k_1+k_2)(x-x_1)}-r_1^{*\uparrow}e^{i(k_1-k_2)(x-x_1)}}{den}-c.c.\right)\\
\fl m_y^\downarrow=-8iq_1q_2(k_3-k_4)\sin\theta\left(\frac{e^{-i(k_1+k_2)(x-x_1)}-r_1^{*\downarrow}e^{-i(k_1-k_2)(x-x_1)}}{den}-c.c.\right)\label{eq:mxmyfin}
\end{eqnarray}

Finally we obtain:
\begin{eqnarray}\label{eq:mx}
\fl m_x=m_x^\uparrow+m_x^\downarrow=&&8q_1q_2(k_3-k_4)\sin\theta\\
&&\times\left(2\left[\frac{e^{-i(k_1+k_2)(x-x_1)}}{den}+c.c.\right]-\left(\left[\frac{r_1^{*\uparrow}}{den}+\frac{r_1^{\downarrow}}{den^*}\right]e^{i(k_1-k_2)(x-x_1)}+c.c.\right)\right) \nonumber\\
\fl m_y=m_y^\uparrow+m_y^\downarrow=&&-8iq_1q_2(k_3-k_4)\sin\theta\left(\left[\frac{r_1^{\uparrow}}{den^*}+\frac{r_1^{*\downarrow}}{den}\right]e^{-i(k_1-k_2)(x-x_1)}-c.c.\right)\label{eq:my}
\end{eqnarray}

By the same way, we evaluate the spin current density for majority and minority spins:
\begin{eqnarray}
\fl J_{x}^{s\uparrow}=-8q_1q_2\frac{\hbar^2}{2m}(k_3-k_4)\sin\theta&&\left(-ik_2\frac{e^{-i(k_1+k_2)(x-x_1)}}{den}+ik_2\frac{r_1^{*\uparrow}e^{i(k_1-k_2)(x-x_1)}}{den}\right.\nonumber\\
&&\left.+ik_1\frac{e^{i(k_1+k_2)(x-x_1)}}{den^*}+ik_1\frac{r_1^{\uparrow}e^{-i(k_1-k_2)(x-x_1)}}{den^*}\right)
\end{eqnarray}

\begin{eqnarray}
\fl J_{x}^{s\downarrow}=-8q_1q_2\frac{\hbar^2}{2m}(k_3-k_4)\sin\theta&&\left(ik_2\frac{e^{i(k_1+k_2)(x-x_1)}}{den^*}+ik_2\frac{r_1^{\downarrow}e^{i(k_1-k_2)(x-x_1)}}{den^*}\right. \nonumber\\
&&\left.-ik_1\frac{e^{-i(k_1+k_2)(x-x_1)}}{den}+ik_1\frac{r_1^{*\downarrow}e^{-i(k_1-k_2)(x-x_1)}}{den}\right)
\end{eqnarray}

\begin{eqnarray}
\fl J_{y}^{s\uparrow}=-8q_1q_2\frac{\hbar^2}{2m}(k_3-k_4)\sin\theta&&\left(ik_2\frac{e^{-i(k_1+k_2)(x-x_1)}}{den}-ik_2\frac{r_1^{*\uparrow}e^{i(k_1-k_2)(x-x_1)}}{den}\right. \nonumber\\
&&\left.+ik_1\frac{e^{i(k_1+k_2)(x-x_1)}}{den^*}+ik_1\frac{r_1^{\uparrow}e^{-i(k_1-k_2)(x-x_1)}}{den^*}\right)
\end{eqnarray}

\begin{eqnarray}
\fl J_{y}^{s\downarrow}=-8q_1q_2\frac{\hbar^2}{2m}(k_3-k_4)\sin\theta&&\left(-ik_2\frac{e^{i(k_1+k_2)(x-x_1)}}{den^*}-ik_2\frac{r_1^{\downarrow}e^{i(k_1-k_2)(x-x_1)}}{den^*}\right.\nonumber\\
&&\left.-ik_1\frac{e^{-i(k_1+k_2)(x-x_1)}}{den}+ik_1\frac{r_1^{*\downarrow}e^{-i(k_1-k_2)(x-x_1)}}{den}\right)
\end{eqnarray}

Taking the imaginary (and real) part of the right-hand-side of the above equations, we obtain, similarly to Eqs. \ref{eq:mx} and \ref{eq:my}:

\begin{equation}\label{eq:jx}
\fl J^s_{x}=-8q_1q_2\frac{\hbar^2}{2m}(k_3-k_4)\frac{(k_1+k_2)}{2}\sin\theta\left(\left[\frac{r_1^{*\uparrow}}{den}+\frac{r_1^{\downarrow}}{den^*}\right]e^{i(k_1-k_2)(x-x_1)}+c.c.\right)
\end{equation}
\begin{eqnarray}\label{eq:jy}
\fl J^s_{y}=-i8q_1q_2\frac{\hbar^2}{2m}(k_3-k_4)\sin\theta&&\left(\left[\frac{e^{i(k_1+k_2)(x-x_1)}}{den^*}-c.c.\right](k_1-k_2)\right. \nonumber\\
&&\left.-\frac{(k_1+k_2)}{2}\left(\left[\frac{r_1^{*\uparrow}}{den}+\frac{r_1^{\downarrow}}{den^*}\right]e^{i(k_1-k_2)(x-x_1)}-c.c.\right)\right)
\end{eqnarray}

The divergency then gives:

\begin{equation}\label{eq:djx}
\frac{\partial J^s_{x}}{\partial x}=-8iq_1q_2\frac{\hbar^2}{2m}(k_3-k_4)\sin\theta\frac{k_1^2-k_2^2}{2}\left(\left[\frac{r_1^{*\uparrow}}{den}+\frac{r_1^{\downarrow}}{den^*}\right]e^{i(k_1-k_2)(x-x_1)}-c.c.\right)
\end{equation}

\begin{eqnarray}
\frac{\partial J^s_{y}}{\partial x}=&& 8q_1q_2\frac{\hbar^2}{2m}(k_3-k_4)\sin\theta\frac{k_1^2-k_2^2}{2}\left(2\left[\frac{e^{-i(k_1+k_2)(x-x_1)}}{den^*}+c.c.\right]\right.\nonumber\\
&&\left.-\left(\left[\frac{r_1^{*\uparrow}}{den}+\frac{r_1^{\downarrow}}{den^*}\right]e^{i(k_1-k_2)(x-x_1)}+c.c.\right)\right)\label{eq:djy}
\end{eqnarray}

Setting $J_{sd}=\frac{\hbar^2}{2m}\frac{k_1^2-k_2^2}{2}$, Eqs. \ref{eq:mx}, \ref{eq:my}, \ref{eq:djx} and \ref{eq:djy} give the following relation:

\begin{eqnarray*}
\frac{\partial J^s_{x}}{\partial x}=-J_{sd}m_y\\
\frac{\partial J^s_{y}}{\partial x}=J_{sd}m_x
\end{eqnarray*}

\begin{equation}
\Rightarrow \nabla{\bf J}^s=J_{sd}{\bf M}\times{\bf m}
\end{equation}

Then, the relation \ref{e:spincont4} can be derived analytically in the free electron approach. This relation does not depend on the particular description adopted (Tight-binding or free electron approximation) but emerges from the definition of the considered Hamiltonian itself.

\section{\label{s:results}Results and discussion}

To illustrate the above calculation, we use material parameters adapted to the case of Co/Al$_2$O$_3$/Co structure: the Fermi wavevectors for majority and minority spins are respectively $k_F^{\uparrow}=1.1$ \AA$^{-1}$, $k_F^{\downarrow}=0.6$ \AA$^{-1}$, the barrier height is $U-E_F=1.6$ eV, the effective electron mass within the insulator is $m_{eff}$=0.4 \cite{bratkovsky} and the barrier thickness is $d$=0.6 nm. These parameters have been choosen to fit the experimental I-V characteristics of the magnetic tunnel junctions studied in Ref. \cite{Petit}. In all this section, the magnetizations form an angle $\theta$=90° between them. We will justify this choice in the following.\par

\subsection{Anatomy of spin transport} 

Although spin-dependent tunnelling is a well known process, the description we give here is of great importance to understand the specific characteristics of spin transfer torques in tunnelling transport. In this part, we will consider the linear approximation in which the bias voltage $V_b$ is low enough so that the current is due to Fermi electrons injected from the left electrode. When the electrodes magnetizations are non collinear, the electrons are no more described as pure spin states, but as a mixing between majority and minority states. For example, let us consider one electron from the left reservoir, initially in majority spin state, impinging on the right electrode (see Fig. 2 - step 1). The first reflection (step 2) at the $F_L/I$ interface do not introduce any mixing since the insulator is non magnetic. However, when (the transmitted part of) this electron is reflected or transmitted by the second interface $I/F_R$ (step 3), the resulting state in the right electrode is a mixing between majority and minority states since the quantization axis in the right electrode is different from the quantization axis in the left electrode. Then, the transmitted spin is reoriented and precesses (step 4) around the magnetization of the right electrode. Furthermore, the reflected electron (step 5) is also in a mixed spin state and precesses around the left electrode magnetization. In other words, after transport through the barrier, the electron spin is reflected/transmitted with an angle. This reorientation gives rise to spin transfer torque.\par

\begin{figure}
	\centering
		\includegraphics[width=8cm]{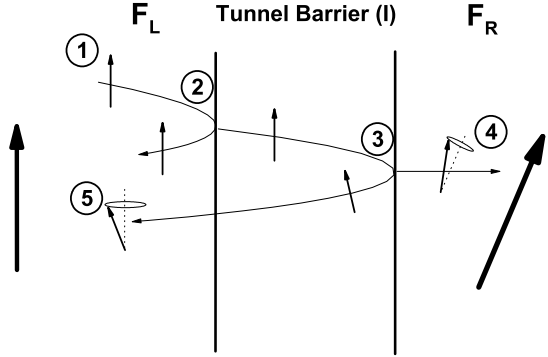}
	\label{fig:trans_noncol}\caption{Schematics of the principle of spin transport in a magnetic trilayer with non collinear electrodes magnetizations. Step 1: the electron spin is polarized along the magnetization of the left electrode. Step 2: After the first reflection/transmission by $F_L/I$ interface the reflected and transmitted parts remain in a pure spin state. Step 3: The reflection/transmission by the second interface $I/F_R$ reorientes the electron spin. Step 4 and 5: The transmitted and reflected spins precess around the local magnetization.}
\end{figure}

Note that there is not reason why the electron spin should remain in the plane of the electrodes magnetizations. We will see that after the reorientation, the electron spin possesses three components in spin space (and so two transverse components).

\subsubsection{Tunnelling transport}

We are first interested in the spin-dependent reflectivity $R^{\sigma(\sigma')}$ and transmittivity $T^{\sigma(\sigma')}$ for electrons at the Fermi energy from the left electrode. Let us consider an electron initially in majority spin state ($\uparrow$). Its wavefunction will be described by a plane wave in the left electrode :

$$\frac{e^{ik_1(x-x_1)}}{\sqrt{k_1}}$$

The mixing between majority and minority spin states can be expressed through mixing reflectivities $R^{\uparrow\uparrow}$ and $R^{\downarrow\uparrow}$ and transmitivities $T^{\uparrow\uparrow}$ et $T^{\uparrow\downarrow}$, so that:

$$R^{\uparrow\uparrow}+R^{\downarrow\uparrow}+T^{\uparrow\uparrow}+T^{\downarrow\uparrow}=1$$

where:

\begin{eqnarray}
&&R^{\uparrow\uparrow}=|r_1^{\uparrow}|^2\\
&&R^{\downarrow\uparrow}=16|\frac{q_1q_2(k_3-k_4)}{m_{eff}^2den}\sin\theta|^2\\
&&T^{\uparrow\uparrow}=|\Psi_L^{\uparrow\left(\uparrow\right)}\frac{d\Psi_L^{*\uparrow\left(\uparrow\right)}}{dx}-\Psi_L^{*\uparrow\left(\uparrow\right)}\frac{d\Psi_L^{\uparrow\left(\uparrow\right)}}{dx}|\\
&&T^{\downarrow\uparrow}=|\Psi_L^{\downarrow\left(\uparrow\right)}\frac{d\Psi_L^{*\downarrow\left(\uparrow\right)}}{dx}-\Psi_L^{*\downarrow\left(\uparrow\right)}\frac{d\Psi_L^{\downarrow\left(\uparrow\right)}}{dx}|
\end{eqnarray}

$\Psi_L^{\sigma(\sigma')}$ is evaluated in the right electrode and given in Appendix. By the same way, we can define the transmittivity and reflectivity of an electron initially in minority spin state. Fig. 3 displays the $\kappa$-dependence of $R$ and $T$ (we omit the superscripts for simplicity), where $\kappa$ is the wavevector component in the plane of the layers.

\begin{figure}
\centering
		\includegraphics[width=10cm]{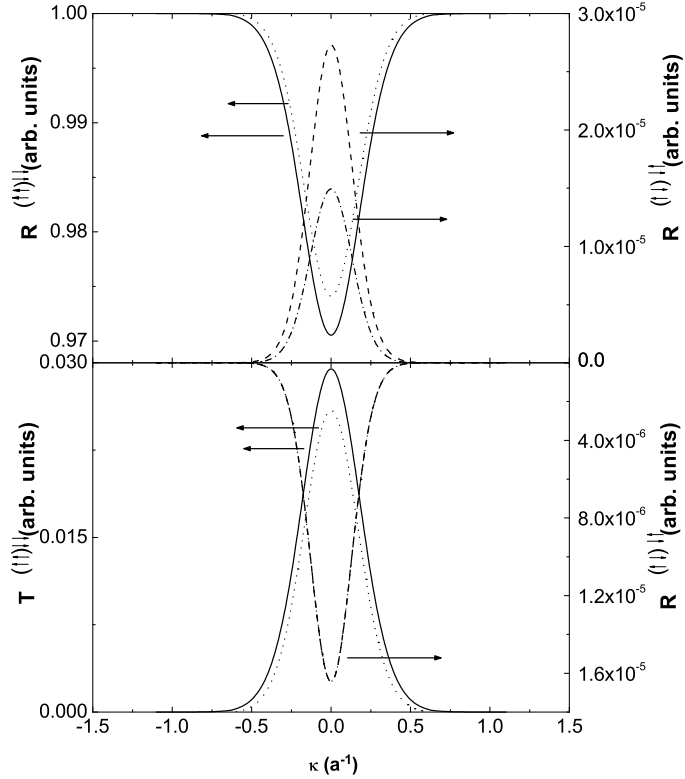}
	\caption{Reflectivity (top panel) and transmittivity (bottom panel) as a function of $\kappa$. The solid and dotted lines represent the spin conserving reflectivity and transmittivity for initially majority and minority spins respectively (left axes); the dashed and dotted-dashed lines represent the mixing reflectivity and transmittivity for initially majority and minority spins respectively (right axes). The applied bias voltage is $V_b=0.1$ V and $\theta$=90°.}
\end{figure}

More than 97\% of the majority and minority spins are reflected conserving their spin projection, whereas less than 3\% are transmitted without spin flip. This reflectivity (transmittivity) reaches a minimum (maximum) at perpendicular incidence and increases (decreases) quickly with $\kappa$. Note that $T^{\uparrow\downarrow}$ and $T^{\downarrow\uparrow}$ are equal due to the particular configuration of the electrodes magnetizations ($\theta$=90°). Thus, after interaction with the barrier, only a very small part of the spin is flipped (the flipped spins have to tunnel through the barrier twice) : less than $2.7\times10^{-3}$\% of the reflected wave has flipped its initial spin. 1.6$\times10^{-3}$\% of the electron spins initially in minority states reverses its spin during reflection.\par
Thus only a very small part of the injected polarized wave is flipped during the tunneling process. However, this does not mean that spin transfer torque is small in MTJ; as a matter of fact, only {\it coherent} mixing states will contribute to transverse spin density, generating spin transfer torque.\par
Finally, we note that only electrons close to the perpendicular incidence contribute significantly to the current. This has important consequences on the impact of quantum interferences on spin transfer.\par

\subsubsection{Spin density and spin transfer torque}

In the linear regime under consideration, the three components of spin density in the left electrode can be described as follows:

\begin{eqnarray}\label{e:m}
\fl m_{+L}^{\uparrow}=\Psi_L^{\uparrow\left(\uparrow\right)}\Psi_L^{*\downarrow\left(\uparrow\right)}=\frac{8q_1q_2(k_3-k_4)\sin\theta}{m_{eff}^2den^*}\left(e^{i(k_1+k_2)(x-x_1)}-r_1^\uparrow e^{-i(k_1-k_2)(x-x_1)}\right)\\
\fl m_{+L}^{\downarrow}=\Psi_L^{\uparrow\left(\downarrow\right)}\Psi_L^{*\downarrow\left(\downarrow\right)}=\frac{8q_1q_2(k_3-k_4)\sin\theta}{m_{eff}^2den}\left(e^{-i(k_1+k_2)(x-x_1)}-r_1^{\downarrow*} e^{-i(k_1-k_2)(x-x_1)}\right)
\end{eqnarray}

\begin{eqnarray}
\fl m_{zL}^\uparrow &=&\Psi_L^{\uparrow\left(\uparrow\right)}\Psi_L^{*\uparrow\left(\uparrow\right)}-\Psi_L^{\downarrow\left(\uparrow\right)}\Psi_L^{*\downarrow\left(\uparrow\right)}\\
&=&\frac{(1+|r_1^\uparrow|^2)}{k_1}-\left|\frac{8q_1q_2\sqrt{k_1}(k_3-k_4)\sin\theta}{m_{eff}^2den}\right|^2-\frac{1}{k_1}\left(r_1^{*\uparrow}e^{2ik_1(x-x_1)}+r_1^{\uparrow}e^{-2ik_1(x-x_1)}\right)\nonumber\\
\fl m_{zL}^\downarrow &=&\Psi_L^{\uparrow\left(\downarrow\right)}\Psi_L^{*\downarrow\left(\uparrow\right)}-\Psi_L^{\downarrow\left(\downarrow\right)}\Psi_L^{*\downarrow\left(\downarrow\right)}\\&=&-\frac{(1+|r_1^\downarrow|^2)}{k_2}+\left|\frac{8q_1q_2\sqrt{k_2}(k_3-k_4)\sin\theta}{m_{eff}^2den}\right|^2+\frac{1}{k_2}\left(r_1^{*\downarrow}e^{2ik_2(x-x_1)}+r_1^{\downarrow}e^{-2ik_2(x-x_1)}\right)\nonumber
\end{eqnarray}

Observing $m_{+L}^{\uparrow\left(\downarrow\right)}$ in Eq. \ref{e:m}, one can distinguish two components: the first one is proportional to $e^{\pm i(k_1+k_2)(x-x_1)}$, and due to the interference between the incident wave with majority (resp. minority) spin and the reflected wave with minority (resp. majority) spin; the second one is proportional to $e^{-i(k_1-k_2)(x-x_1)}$ and due to the interference between the reflected waves with majority and minority spins. We note that the first components of  $m_{+L}^{\uparrow}$ and $m_{+L}^{\downarrow}$ are complex conjugated so that their sum is real. Then, the interference between the incident wave with majority spin and the reflected wave with minority spin does not contribute to STT but only to IEC. STT is then generated by the coherent interferences between reflected electrons with opposite spin projection ($\propto e^{-i(k_1-k_2)(x-x_1)}$).\par
Concerning $m_{zL}$, it is composed of one component proportionnal to $e^{2ik_1(x-x_1)}$, one component proportionnal to $e^{2ik_2(x-x_1)}$ and one constant as a function of $x$. The two formers are due to the interference between wavefunctions in the same spin projection but with opposite propagation direction while the latter is due to interference between wavefunctions in the same spin projection and the same propagation direction.\par

Fig. 4 displays the details of the spin density components $m_x$, $m_y$ et $m_z$ (described in Eq. \ref{e:m}) in the left electrode as a function of $x$, when $V_b=0.1$ V. $m_x$ possesses a quite complex behaviour with two periods of oscillation (the dashed lines show the enveloppe of the curve), whereas $m_y$ is reduced to a single oscillation (The oscillation period $k_1+k_2$ vanishes when suming the contribution of majority and minority spins); $m_z$ oscillates around mean values represented by horizontal dashed lines.\par

\begin{figure}
	\centering
		\includegraphics[width=8cm]{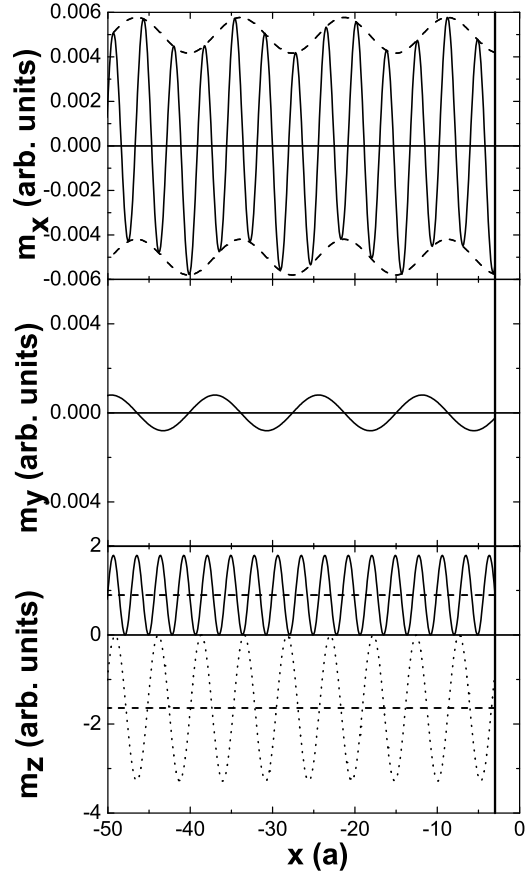}
	\label{fig:torqueleft}\caption{Projections of spin density due to Fermi electrons in perpendicular incidence from the left electrode, as a function of the distance from the interface. Top panel: $m_x$ component of spin density (solid line); the dashed lines are the enveloppes of the curve. Middle panel: $m_y$ component of spin density. Bottom panel: $m_z$ component of spin density due to initially majority (solid line) and minority (dotted line) spin projection; the dashed lines are the mean values of the oscillations. The applied bias voltage is $V_b=0.1$ V. The vertical line on the right is the interface between the left electrode and the tunnel barrier.}
\end{figure}

Note that the conservative part of IEC is only proprotionnal to $e^{-i(k_1+k_2)(x-x_1)}$. But at non zero bias, the dissipative part of IEC is proportionnal to both $e^{-i(k_1+k_2)(x-x_1)}$ and  $e^{-i(k_1-k_2)(x-x_1)}$. Then, at non zero bias, the electrons will not precess circularly around the background magnetization, but will present a more complex structure.\par
Following the previous discussion about spin reorientation (see Fig. 2), it is possible to deduce the angles at which the electron spin is reflected by the barrier. We define the azimuthal angle azimuthal $\eta$ and the polar angle $\phi$ as indicated in the insert of Fig. 5:

\begin{eqnarray}
&&\eta=\arctan \frac{m^c_y}{m^c_x}\\
&&\phi=\arccos \frac{m^c_z}{\sqrt{m^{c2}_x+m^{c2}_y+m^{c2}_z}}
\end{eqnarray}

In definition of the vector $\overrightarrow{m}^c$, we only considered the coherent interferences between plane wave propagating in opposite direction ($\propto e^{\pm i(k_{1}-k_{2})}$ and constant component of $m_z$), as discussed above. Fig. 5 shows the $\kappa$-dependence of these angles at the interface $F_L/I$ (x=-3 \AA) for an electron spin initially in majority state and for different barrier thicknesses (top panel) and heights (bottom panel). The azimuthal angle $\eta$ varies between -64° to +77° while the polar angle $\phi$ remains very small (less than 0.2°), which means that the electron spin stays very close to the quantization axis, as discussed above. At $\kappa=0.6$ \AA$^{-1}$ (corresponding to $k_F^{\downarrow}$), $\eta=0$ which indicates that the effective spin density lies in the plane of the magnetizations $\left(\overrightarrow{M}_L,\overrightarrow{M}_R\right)$. Finally, the polar angle does not vary with the distance, which means that the reflected electron spin precesses around $O_z$ with a small angle $\phi$. A "Bulk" spin transfer only occurs under the interferences of all the reflected electrons.\par

The strong dependence of $\eta$ as a function of the in-plane wavevector $\kappa$, together with the dominant contribution of nearly perpendicularly incident electrons (see Fig. 3), implies that the effective electron spin, resulting from the averaging over all the incident electrons, possesses an important out-of-plane component. In other words, the effect of the spin-dependent tunneling is to strongly enhance the dissipative IEC component of the spin torque, compared to metallic spin valves. As a matter of fact, in SV the whole Fermi surface contributes to the spin transport so that the effective angle $\eta$ is very small \cite{stiles02}: the dissipative IEC is thus negligible.\par

Note that increasing the thickness of the barrier only weak influence on $\eta$ and strongly decreases the amplitude of $\phi$ (the mixing reflection decreases since the barrier thickness increases, then reducing the transverse spin density). Furthermore, when increasing the barrier height, both amplitudes of the angles $\phi$ and $\eta$ decreases near the perpendicular incidence. These results are consistent with the reduction of spin mixing when decreasing the barrier transmittivity.\par

\begin{figure}
	\centering
		\includegraphics[width=10cm]{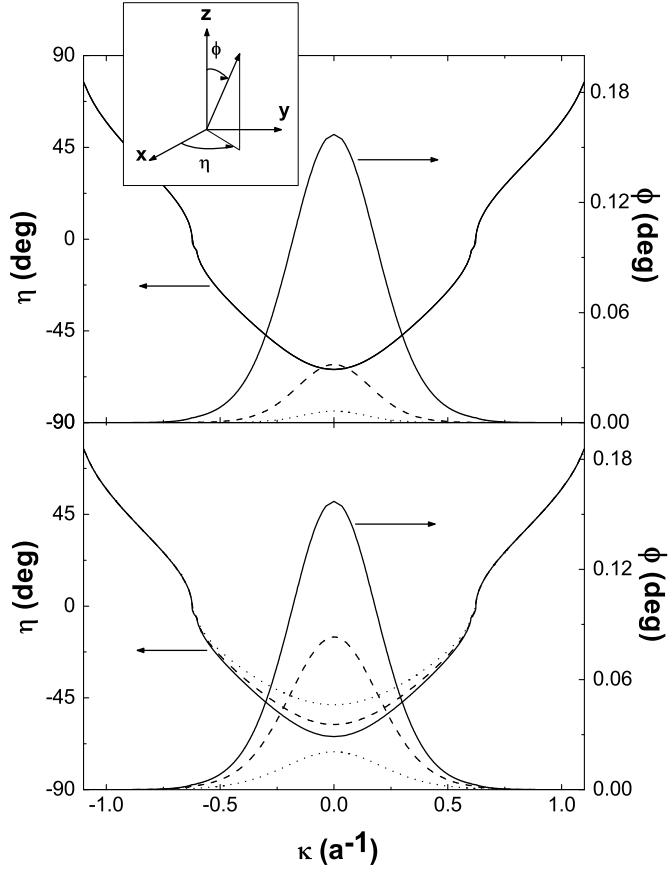}
	\label{fig:anglespin}\caption{$\kappa$-dependence of the reflection angles for an electron spin at Fermi energy, initially in majority spin state. Top panel: the barrier thickness is set to d=0.6 nm (solid line), d=0.8 nm (dashed line) and d=1 nm (dotted line); $U-E_F$=1.6 eV. Bottom panel: the barrier height is set to $U-E_F$=1.6 eV (solid line), $U-E_F$=2 eV (dashed line) and $U-E_F$=3 eV (dotted line); d=0.6 nm. Insert: Definition of the angles $\phi$ and $\eta$. The applied bias voltage is $V_b=0.1$ V and $\theta$=90°.}
\end{figure}

Fig. 6 shows the dependence of the angles as a function of the {\it s-d} exchange constant $J_{sd}$ for perpendicular incidence $\kappa=0$. Quite intuitively, the precession angle $\phi$ increases with $J_{sd}$ whereas the initial azimuthal angle $\eta$ decreases in absolute value with $J_{sd}$. The spin-filtering effect (the selection between majority and minority spin during the reflection process) increases with $J_{sd}$ so that $\overrightarrow{m}_c$ gets closer to the plane of the magnetizations.

\begin{figure}
	\centering
		\includegraphics[width=10cm]{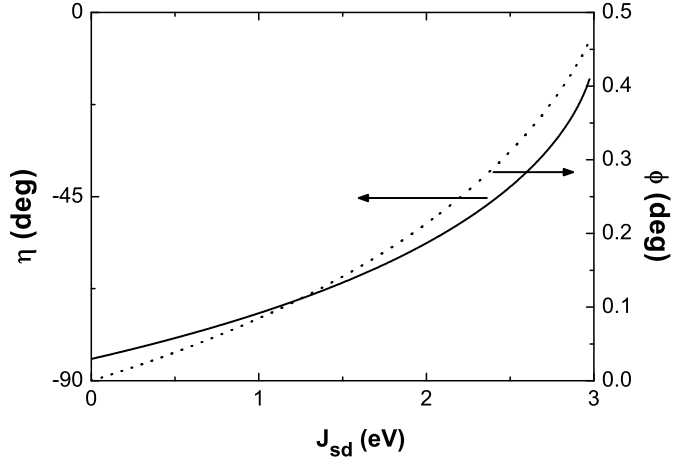}
	\label{fig:anglespin_Jsd}\caption{Reflection angles as a function of the {\it s-d} exchange constant, for a Fermi electron initially in majority spin state. The parameters are the same as in Fig. 5.}
\end{figure}

\subsection{Spin Transfer Torques}

We now take into account all the electrons in the calculations (from the left and the right electrodes). Fig. 7 shows STT and IEC as a function of the angle $\theta$ between the electrodes magnetizations, at $V_b=0$ and $V_b=0.1$ V. It clearly appears that IEC and STT are proportionnal to $\sin\theta$ (the deviation from $\sin\theta$ is minor than 10$^{-4}$). This dependence is strongly different from what was predicted in metallic spin valves \cite{autres,jpcm,slonc02} and has been attributed \cite{slonc05} to the single-electron nature of tunneling. As a matter of fact, because of the important height of the tunnel barrier, all the potential drop occurs inside the insulator and spin accumulation (i.e. the feedback of the current-induced longitudinal spin density on the spin current) is negligible. In this case, the angular dependence of torque is determined by the angular dependence of the transmition matrix, as discussed in Ref. \cite{slonc05} and yields a sine shape. In the following, we will estimate the spin density for $\theta=\pi/2$.\par
Note that, at zero bias, interlayer exchange coupling is still non-zero, contrary to spin transfer torque. The conservative part of IEC (IEC at zero bias) comes from the contribution of electrons located under the Fermi level. At zero bias, the currents from left and right electrodes are equal, but the electron propagation still corresponds to the scheme shown in Fig. 2: the mixing between majority and minority states induces a transverse component in the spin density.\par

\begin{figure}
	\centering
		\includegraphics[width=8cm]{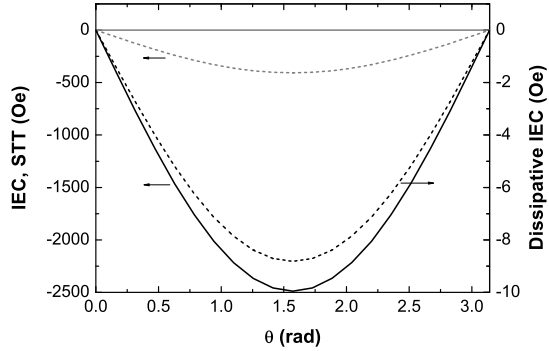}
	\label{fig:STT_IEC_theta}\caption{Angular dependence of spin transfer (grey) and interlayer exchange coupling (black): conservative part (at zero bias - solid lines) and dissipative parts (bias dependent part - dashed lines). The dissipative parts are calculated at $V_b=0.1$ V.}
\end{figure}

Fig. 8 displays the two components of transverse spin density as a function of the location in the left electrode. The interference process between polarized electrons yields a damped oscillation of IEC as presented in Fig. 8(a). We can distinguish two periods of oscillation $T_1=2\pi/\left(k_F^\uparrow-k_F^\downarrow\right)$ and $T_2=2\pi/\left(k_F^\uparrow+k_F^\downarrow\right)$ whereas at zero bias, only $T_2$ appears (see inset of Fig. 8(a)). This can be easily understood by considering electrons from left and right electrodes. Transverse spin density in the {\it left} electrode due to electrons from the {\it right} electrode is:

\begin{eqnarray}\label{e:m2}
m_{+R}^{\uparrow} & =& \Psi_R^{\uparrow\left(\uparrow\right)}\Psi_R^{*\downarrow\left(\uparrow\right)}\\
&=&\left|\frac{8\sqrt{q_1q_2}}{m_{eff}^2den}\right|^2\frac{\sin\theta}{2}k_3\Psi(q_1,k_2,q_2,k_4)\Psi^*(q_1,k_1,q_2,k_4)e^{-i(k_1-k_2)(x-x_1)}\\
m_{+R}^{\downarrow} & =& \Psi_R^{\uparrow\left(\downarrow\right)}\Psi_R^{*\downarrow\left(\downarrow\right)}\\
&=&\left|\frac{8\sqrt{q_1q_2}}{m_{eff}^2den}\right|^2\frac{\sin\theta}{2}k_4\Psi(q_1,k_2,q_2,k_3)\Psi^*(q_1,k_1,q_2,k_3)e^{-i(k_1-k_2)(x-x_1)}
\end{eqnarray}

It is now possible to show that in the general expression of transverse spin density
$$m_+=m_{+L}^{\uparrow}+m_{+L}^{\downarrow}+m_{+R}^{\uparrow}+m_{+R}^{\downarrow}$$
the terms proportional to $e^{-i(k_1-k_2)(x-x_1)}$ vanish at zero bias and $m_+$ reduces to terms proportional to $e^{\pm i(k_1+k_2)(x-x_1)}$. Furthermore, these last terms only give a real component since, as discussed above, the majority and minority components of $m_y$ compensate each other. Consequently, at zero bias, only the conservative part of interlayer exchange coupling exists, due to the interference between incident and reflected electrons with opposite spin projection. But when the bias voltage is non zero, the transport becomes asymetric and the terms proportional to $e^{-i(k_1-k_2)(x-x_1)}$ do not compensate each other anymore and lead to two periods of oscillations as shown in Fig. 8(a).\par
 
\begin{figure}
	\centering
		\includegraphics[width=8cm]{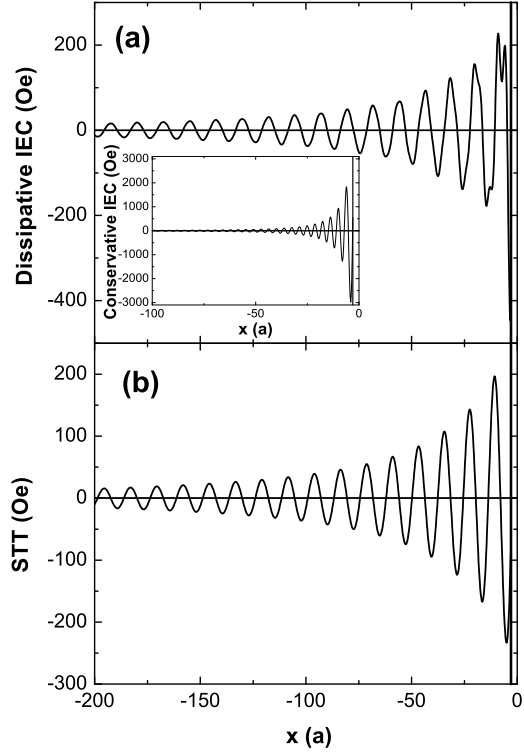}
	\label{fig:torques_z}\caption{Total spin density as a function of the location in the left electrode: a) Current-induced interlayer exchange coupling - inset: Interlayer exchange coupling at zero bias voltage; b) Spin transfer torque. These quantities are calculated at $V_b=0.1$ V.}
\end{figure}

Spin transfer torque, proportional to $m_y$, only exits at non zero bias and possesses only one period of oscillation $T_1$ (see Fig. 8(b)). It is worthy to note that the transverse components of spin density is damped by 50\% within the first nanometers, and that the amplitude of IEC is of the same order than STT. This decay lenght is very large compared to previous theoretical predictions \cite{stiles02,slonc02} and experimental investigations on SV \cite{urzh}. As a matter of fact, the ballistic assumption holds for distance smaller than the mean free path ($\approx 5$ nm in Co). In realistic devices, spin diffusion should increase the decay of STT and IEC. Another source of this difference compared to metallic SV is the fact that we consider perfect interfaces and no defaults in the barrier. First principle studies of realistic Co/Cu interfaces \cite{zwier} showed that the mismatch of the electronic structure at the interface strongly reduces the transverse component of spin density. In MTJ, the non spherical nature of the spin-dependent Fermi surface \cite{belashchenko,butler} should also dramatically alter the transverse spin density. This could also explain the fact that the amplitude of spin torque is two orders of magnitude higher than in experiments.\par
Another characteristic specific to MTJ is that in our calculation we find that dissipative IEC is of the same order of magnitude than STT. This is coherent with the theoretical results of Theodonnis et al. \cite{theo} as well as with the experimental studies of Petit et al. \cite{Petit}. This can be attributed to the high $\kappa$-selection due to the tunneling transport. We previously found that the contribution to torque strongly decrease with $\kappa$ (see Figs. 3 and 5) so that only electrons with small $\kappa$ strongly contribute to spin torque. In this case, the averaging of torques (and specifically IEC) will be less destructive than in metallic spin valves where all the Fermi surface is involved in the quantum interferences.

\begin{figure}
	\centering
		\includegraphics[width=10cm]{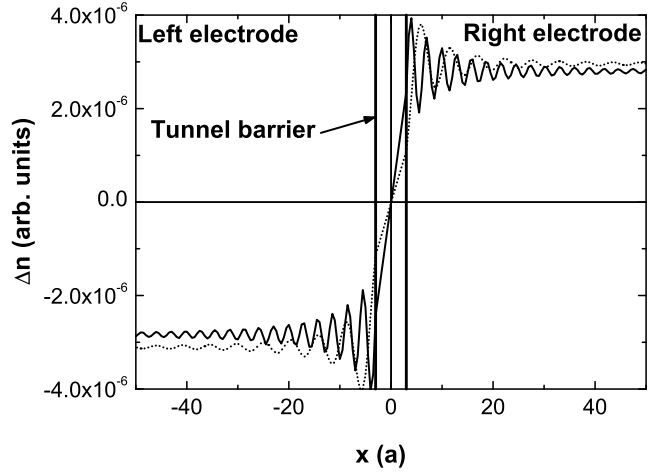}
	\label{fig:mz_z}\caption{Out-of-equilibrium longitudinal spin density along the magnetic tunnel junction for majority (solid line) and minority (dotted line) electron spin projections. The bias voltage is $V_b=0.1$ V.}
\end{figure}

Finally, Fig. 9 shows the out-of-equilibrium longitudinal spin density $\Delta n$ defined as $\Delta n^{\uparrow\left(\downarrow\right)}=n^{\uparrow\left(\downarrow\right)}(V_b=0.1)-n^{\uparrow\left(\downarrow\right)}(V_b=0)$. $\Delta n$ oscillates and asymptotically reaches a non zero value. This means that when the bias voltage is turned on, a non equilibrium spin accumulation builds up. However, this effective spin accumulation is very small ($\Delta n^{\uparrow}-\Delta n^{\downarrow}\approx 10^{-7}$ electron/atom) and cannot influence spin current building. Then, neglecting the role of longitudinal spin accumulation (spin density) in MTJ is justified.\par

\subsection{Bias dependence}

The bias dependence of STT and IEC in MTJ also presents strong differences with SV. We first calculate the total spin torque exerted on the left electrode. Following the definition of Ref. \cite{slonc} and Ref. \cite{theo}, the total torque is:
\begin{equation}\label{eq:js}
	\overrightarrow{T}_{total}=\int_{x_1}^{-\infty}-\nabla \bm{J}^s dx=\bm{J}^s(x_1)
\end{equation}

Fig. 10 displays the total interlayer exchange coupling (a) and spin transfer torque (b) as a function of the applied bias voltage, for different values of the {\it s-d} exchange parameter $J_{sd}$. Our results are consistent with Theodonnis et al. \cite{theo}. The dissipative IEC is quadratic whereas STT is a combination between linear and quadratic bias dependence. In Ref. \cite{theo}, the authors proposed a general formula, derived from Slonczewski circuit theory \cite{slonc02}, linking total spin transfer torque with interfacial spin current densities \cite{theo}:

\begin{equation}\label{eq:js2}
	T_{||}=\frac{J^s_{AP}-J^s_{P}}{2}
\end{equation}

where $J^s_{AP(P)}$ are interfacial spin current densities when the electrodes magnetizations are antiparallel and parallel respectively (see the definition in Ref. \cite{theo}). The authors claimed that this relation should hold for any electronic structure, so any transport description. As a matter of fact, the top inset of Fig. 10(b) shows STT calculated using Eq. \ref{eq:js} (solid line) and Eq. \ref{eq:js2} (symbols). It shows very good agreement between the two members of Eq. \ref{eq:js2}.

\begin{figure}
		\includegraphics[width=16cm]{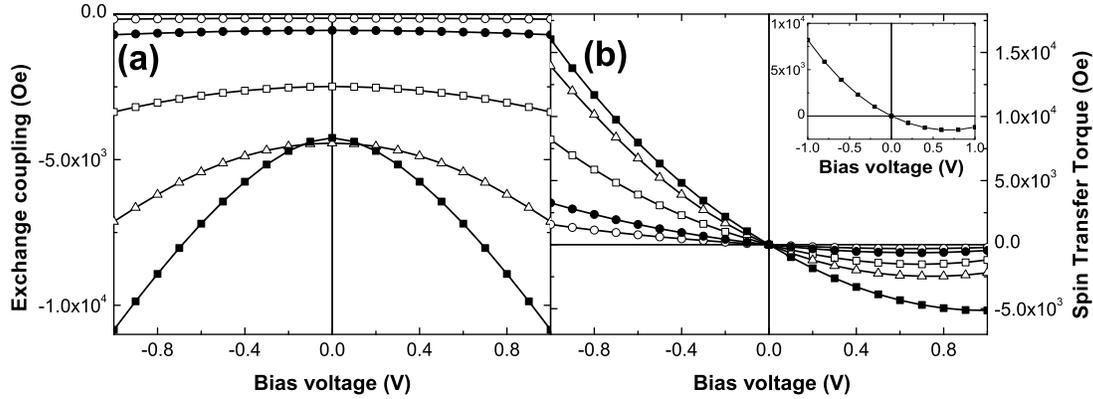}
	\label{fig:STT_V_Jsd}\caption{Bias dependence of interlayer exchange coupling (a) and spin transfer torque (b) for different values of {\it s-d} coupling: $J_{sd}=0.38$ eV (open circles), $J_{sd}=0.76$ eV (filled circles), $J_{sd}=1.62$ eV (open squares), $J_{sd}=2.29$ eV (open triangles), $J_{sd}=2.97$ eV (filled squares). Top inset: Bias dependence of STT for $J_{sd}=1.62$ eV; the solid line was calculated following the usual way and the symbols were calculated using Eq. \ref{eq:js2}.}
\end{figure}

Experimental studies by Cornell's group \cite{Sankey,fuchs} demonstrated a linear variation of spin transfer torque as a function of the applied bias voltage. This linear variation is also usually assumed in interpreting excitations studies \cite{kubota,Petit}. Moreover, the very recent article of Sankey et al. \cite{Sankey} seems to confirm the fact that the dissipative exchange coupling is quadratic as a function of the bias voltage. Finally, note that a change of sign of spin transfer torque at high positive bias voltage is expected, consistently with Ref. \cite{theo}. The STT change of sign should be observed in MTJ with low enough barrier height and high breakdown voltage (MgO seems a good candidate). Nevertheless, more technological development is needed to fabricate such junctions.\par
Eq. \ref{eq:js} assumes that all the transverse spin density has been absorbed in the free layer. However, in very thin free layer, one can expected that transverse spin density is not fully absorbed when leaving the free layer. In this case, one should consider that the free layer is finite. Fig. 11 displays the bias dependence of IEC and STT for different integration depths $t$ (namely, different layer thicknesses): 

\begin{equation}\label{eq:js23}
	\overrightarrow{T}_{partial}=\int_{x_1}^{x_1-t}-\nabla \bm{J}^s dx=\bm{J}^s(x_1)-\bm{J}^s(x_1-t)
\end{equation}

The dependence can change drastically and IEC can even change its sign (note that STT keeps its general shape). These dependences are strongly affected by the tunnel barrier characteristics and one should to be careful in the analysis of bias dependence.\par

\begin{figure}
		\includegraphics[width=16cm]{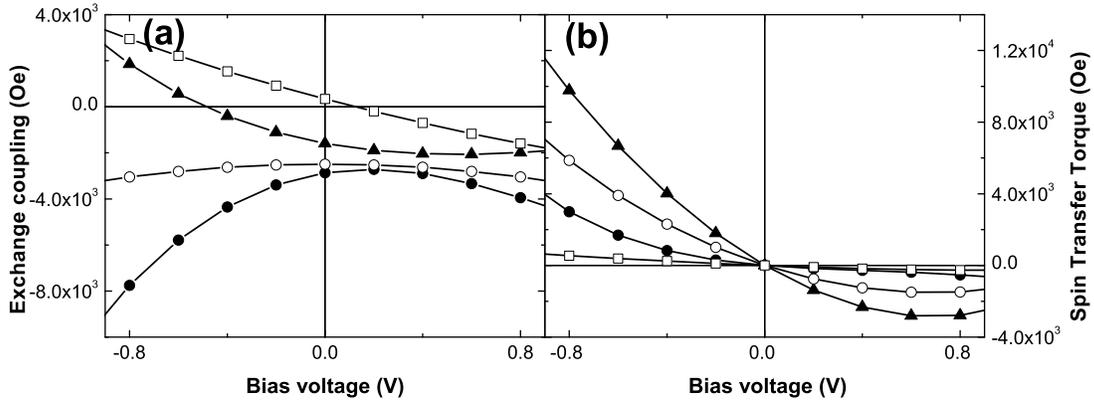}
	\caption{Bias dependence of interlayer exchange coupling (a) and spin transfer torque (b) for $J_{sd}=1.62$ eV and different values integration depth: $t=0$ \AA (open squares), $t=4$ \AA (filled triangles), $t=10$ \AA (filled circles), $t=\infty$ \AA (open circles).}\label{fig:fig11}
\end{figure}

\subsection{From weak ferromagnetic to half-metallic tunnel junction}

To conclude this article, we study the dependence of the total spin transfer torque and interlayer exchange coupling as a function of the energy of the bottom of the minority electrons conduction band $\epsilon^{\downarrow}$, as indicated in Fig. \ref{fig:fig11}. This energy is defined from the Fermi energy as:
\begin{equation}
	\epsilon^{\downarrow}=E_F-E_c^{\downarrow}=-\frac{\hbar^2k_F^{\downarrow2}}{2m}
\end{equation}
where $E_c^{\downarrow}$ is the absolute energy of the bottom of the conduction band. In the present study, we vary $\epsilon^{\downarrow}$, keeping $\epsilon^{\uparrow}$ and $E_F$ constant. When $\epsilon^{\downarrow}$ is close to $\epsilon^{\uparrow}$, $k_F^{\uparrow}\approx k_F^{\downarrow}$, the metallic electrodes loose their ferromagnetic nature. For $\epsilon^{\downarrow}\approx 0$, the Fermi wavevector for minority electrons becomes smaller and the current polarization is strongly enhanced. In this case, we expect an important spin transfer torque. When $\epsilon^{\downarrow}>0$, $k_F^{\downarrow}$ becomes imaginary and the electrodes behave like a tunnel barrier for minority spins. Increasing $\epsilon^{\downarrow}$ increases the evanescent decay of minority wavefunctions in the electrodes. Then, the product $<\Psi^{*\uparrow}\Psi^{\downarrow}>$ still exists so that spin torque is non zero and decrease exponentially from the interface.\par

Fig. 12 shows the amplitude of total STT and current-induced IEC in the three different regimes: weak ferromagnetic electrodes (WFM), strong ferromagnetic electrodes (SFM) and half-metallic electrodes (HM). As expected, in ferromagnetic regime, STT and dissipative IEC increase until $\epsilon^{\downarrow}=0$ (vertical line). When $\epsilon^{\downarrow}$ becomes positive, the bottom of the conduction band of minority electrons lies above the Fermi level: no minority electrons can propagate because only evanescent states exist near the interfaces for this spin projection. However, STT and dissipative IEC do not vanish but reach a plateau which slowly decreases to zero when increasing $J_{sd}$ (not shown).

\begin{figure}
	\centering
		\includegraphics{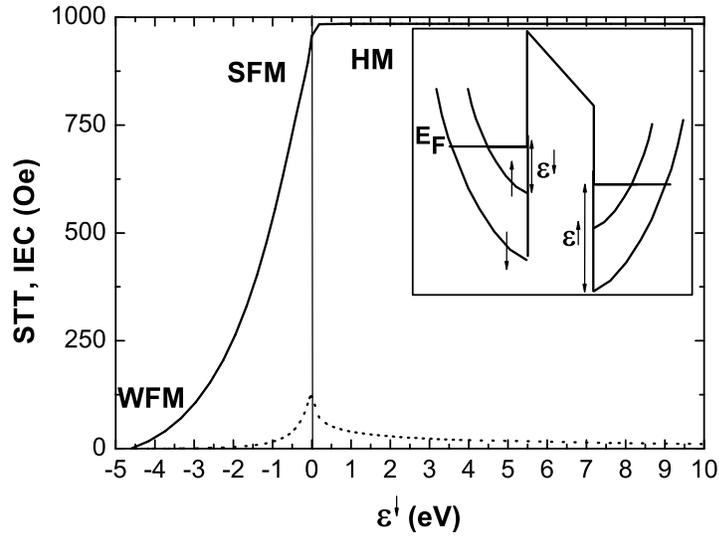}
	\label{fig:fig12}\caption{Spin transfer torque (solid line) and dissipative interlayer exchange coupling (dotted line) as a function of {\it s-d} exchange coupling. The vertical line shows the limit between ferromagnetic (weak ferromagnetic -WFM- and strong ferromagnetic -SFM-) regime and half-metallic regime.}
\end{figure}

To understand this behaviour, we calculated the spatial dependence of the transverse spin density in the free layer. Fig. 13 shows the transverse spin density in a usual ferromagnet, $\epsilon^\downarrow=-1.37$ eV (which corresponds to $J_{sd}=1.62$ eV), as a function of the distance from the interface in the left electrode. The oscillation possesses the same characterisics than discussed above and we observe that the transverse spin density is damped far from the interface. When decreasing $\epsilon^\downarrow$, the interfacial spin density increases, due to strong spin filtering at the interface (strong spin-dependent selection), as shown on Fig. 14.

\begin{figure}
	\centering
		\includegraphics{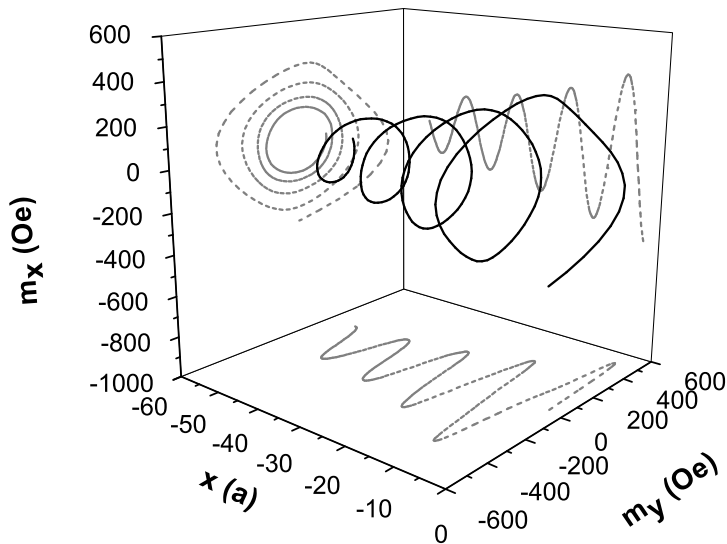}
	\label{fig:fig13}\caption{Transfer spin density (black line) as a function of the distance in the left ferromagnetic electrode in a usual ferromagnetic regime. We set $\epsilon^\downarrow=-1.37$ eV and $V_b=0.1$ V.}
\end{figure}

\begin{figure}
	\centering
		\includegraphics{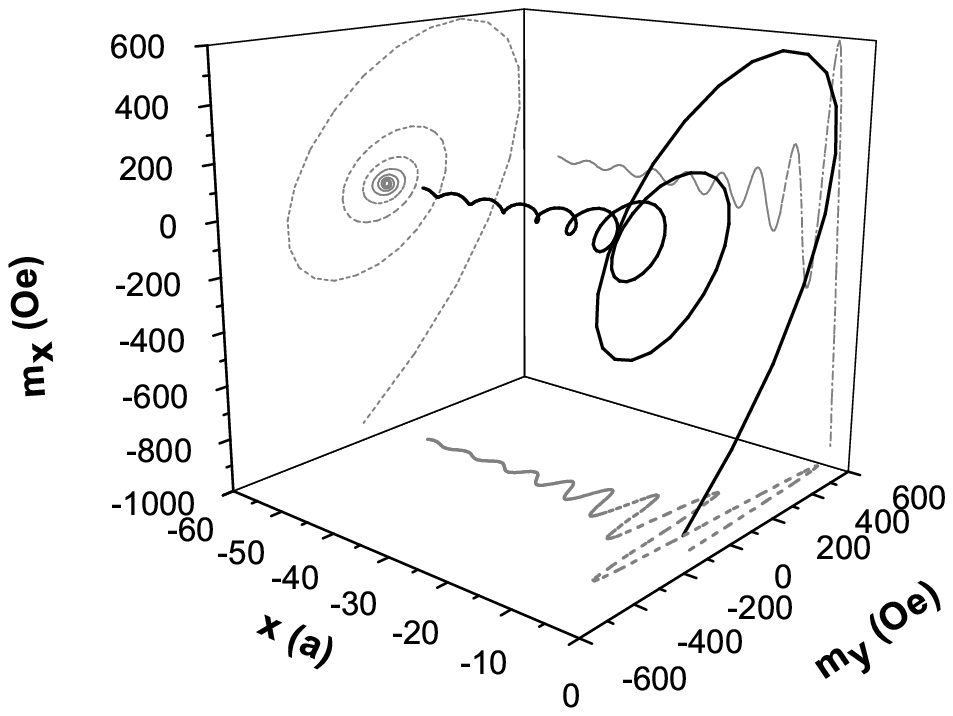}
	\label{fig:fig14}\caption{Transfer spin density (black line) as a function of the distance in the left ferromagnetic electrode in a strong ferromagnetic regime. We set $\epsilon^\downarrow=-0.38$ eV and $V_b=0.1$ V.}
\end{figure}

But when $\epsilon^\downarrow$ changes its sign, only majority electrons can propagate and the transverse spin density is (see Eqs. \ref{eq:mxmy}-\ref{eq:mxmyfin}):
\begin{eqnarray}
\fl m_x^\uparrow=16q_1q_2\sin\theta\ \Re\{(k_3-k_4)\left(\frac{e^{-i(k_1+k_2)(x-x_1)}-r_1^{*\uparrow}e^{i(k_1-k_2)(x-x_1)}}{den}\right)\}\\
\fl m_y^\uparrow=-16q_1q_2\sin\theta\ \Im\{(k_3-k_4)\left(\frac{e^{-i(k_1+k_2)(x-x_1)}-r_1^{*\uparrow}e^{i(k_1-k_2)(x-x_1)}}{den}\right)\}
\end{eqnarray}
Considering Fermi electrons at perpendicular indidence, very small bias voltage ($eV\approx0$) and imaginary minority electron spin wavevector, $k_{2(4)}=ik$, we obtain straightforwardly:

\begin{eqnarray}
\fl m_x^\uparrow=16q_1q_2 e^{k(x-x_1)}\sin\theta\ \Re\{(k_3-ik)\left(\frac{e^{-ik_1(x-x_1)}-r_1^{*\uparrow}e^{ik_1(x-x_1)}}{den}\right)\}\\
\fl m_y^\uparrow=-16q_1q_2e^{k(x-x_1)}\sin\theta\ \Im\{(k_3-ik)\left(\frac{e^{-ik_1(x-x_1)}-r_1^{*\uparrow}e^{ik_1(x-x_1)}}{den}\right)\}
\end{eqnarray}

The transverse spin density is a product between oscillating function of $k_1$ and exponentially decaying function of $k$. Fig. 15 shows the spatial evolution of the transverse spin density in the case of a half-metallic tunnel junction. All the oscillations are damped very quickly so that the only important contribution to torque comes from the interface. Contrary to usual MTJ (where both bulk averaging due to spatial interferences and interfacial spin reorientation contribute to spin torque), in a strong half-metallic tunnel junction all the torque comes from spin reorientation due to spin-dependent reflection. In this last case, the contribution of the spatial averaging between all the impinging electrons ($\kappa$-summation) is reduced compared to interfacial spin transfer.

\begin{figure}
	\centering
		\includegraphics{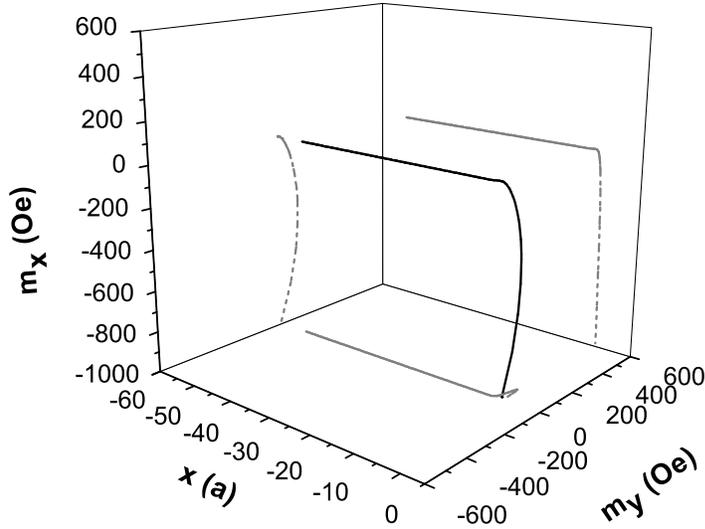}
	\label{fig:fig15}\caption{Transfer spin density (black line) as a function of the distance in the left ferromagnetic electrode in half-metallic regime. We set $\epsilon^\downarrow=19$ eV and $V_b=0.1$ V.}
\end{figure}

\section{\label{s:concl}Conclusion}

A free-electron {\it s-d} model has been proposed to analyze spin transfer effects in magnetic tunnel junctions with amorphous barrier and non collinear electrode magnetizations. We first studied the anatomy of spin transport in such MTJ, showing that only a small part of the current undergoes spin-flipping due to the non collinearity of the electrode magnetizations. This corresponds to only a small deviation of the reflected spin from the local magnetization. Nevertheless, we showed that this small amount of precessing spin gives rise to an important transverse spin density leading to spin torque.\par

We also showed that the tunnel barrier acts like an incidence filter which increases the contribution of the electrons impinging with angle close to the perpendicular incidence. This $\kappa$-selection is at the origin of an important IEC, contrary to what is observed experimentally in metallic spin valves. The ballistic transport dominating the tunnel transport in MTJ is expected to induce large oscillations of STT and IEC as a function of the distance from the interface. If the oscillation period is large compared to the exchange length and one will observe a twist of the magnetization in the thickness of the layer. Otherwise, if the oscillation period is short compared to the exchange length, one will observe the torque integrated over the layer thickness.\par

The bias dependence of spin transfer torque shows a strong asymmetry and a change of sign at positive bias voltage. This results is coherent with tight-binding calculations \cite{theo}. However, we saw that this model is strongly limited to small bias voltage because of the simplicity of the adopted band structure.\par

Finally, we analyzed STT and IEC when varying the {\it s-d} exchange coupling and we demonstated that the torque still exists in MTJ composed of half-metallic electrodes, due to spin-dependent reflections. However, for infinite half-metallic MTJ (for infinite {\it s-d} coupling), it is shown that STT and dissipative IEC vanishes to zero.\par

Furthermore, several numerical studies have shown that, even in amorphous barrier, the interfaces composition and specially the presence of interfacial oxygen have a very deep influence on the spin polarization and thus on TMR and STT \cite{belashchenko}. The recent development of MgO-based MTJ in spin transfer studies reduced the interest in amorphous barriers. However, amorphous barriers have the ability to present a simple physical framework which can constitute a basis to understand spin transfer in MTJ. Nevertheless, because of its more complex band structure and spin-filtering effect associated with the symmetry of wavefunctions, microscopic analysis of spin transfer in MgO-based MTJ would present exciting fundamental characteristics even on spin transfer effects \cite{butler}.

\ack
The work and results reported in this publication were obtained with research funding from RTN SPINSWITCH and from Russian Fundings for Basic Research 07-02-00918-a. The views expressed are solely those of the authors, and the other Contractors and/or the European Community cannot be held for any use that may be made of the information contained herein.

\section*{Appendix: Spin-dependent wave-functions in a clean MTJ\label{app}}

In this appendix, we give the analytical formulae for the spin dependent wavefunctions in the MTJ. Some functions which will be used in the description of this wavefunctions are first defined:

\begin{eqnarray*}
\fl q_0^2=\frac{2m}{\hbar^2}\left(U-E_F\right)\\
\fl q(x)=\sqrt{q_0^2-\frac{2m}{\hbar^2}\left(\frac{x-x_1}{x_2-x_1}eV-\epsilon\right)+\kappa^2}\\
\fl q_{1}=q(x_1)\\
\fl q_{2}=q(x_2)\\
\fl k_{1(2)}=\sqrt{\left(k_F^{\uparrow\left(\downarrow\right)}\right)^2-\frac{2m}{\hbar^2}\epsilon-\kappa^2}\\
\fl k_{3(4)}=\sqrt{\left(k_F^{\uparrow\left(\downarrow\right)}\right)^2-\frac{2m}{\hbar^2}\left(\epsilon-eV\right)-\kappa^2}\\
\fl E(x_i,x_j)=\exp{\int_{x_i}^{x_j}q(x)dx}\\
\fl E_n=E(x_1,x_2)
\end{eqnarray*}
where $E_F$ is the Fermi energy, U is the height of the barrier, V is the bias voltage and $\epsilon=E_F-E$, $E$ being the energy of tunnelling electron. We define:

\begin{eqnarray*}
\fl \Psi(q_1,k_i,q_2,k_j)=E_n(q_1-ik_i)(q_2-ik_j)-E_n^{-1}(q_1+ik_i)(q_2+ik_j)\\
\fl \phi(q_1,k_i,q_2,k_j)=E_n(q_1+ik_i)(q_2-ik_j)-E_n^{-1}(q_1-ik_i)(q_2+ik_j)\\
\fl den=\Psi(q_1,k_1,q_2,k_3)\Psi(q_1,k_2,q_2,k_4)(1+\cos\theta)+\Psi(q_1,k_2,q_2,k_3)\Psi(q_1,k_1,q_2,k_4)(1-\cos\theta)\\
\fl r_1^{\uparrow}=\frac{1}{den}\left[\phi(q_1,k_1,q_2,k_3)\Psi(q_1,k_2,q_2,k_4)(1+\cos\theta)+\phi(q_1,k_1,q_2,k_4)\Psi(q_1,k_2,q_2,k_3)(1-\cos\theta)\right]\\
\fl r_3^{\uparrow}=\frac{1}{den}\left[\phi(q_2,k_3,q_1,k_1)\Psi(q_1,k_2,q_2,k_4)(1+\cos\theta)+\phi(q_2,k_3,q_1,k_2)\Psi(q_1,k_1,q_2,k_4)(1-\cos\theta)\right]\\
\end{eqnarray*}

Electrons initially in the left electrode have the following wavefunctions along the structure :

\begin{eqnarray*}
\fl 
\Psi_L^{\uparrow\left(\uparrow\right)}\left(-\infty<x<x_1\right)&&=\frac{1}{\sqrt{k_1}}\left[e^{ik_1\left(x-x_1\right)}-r_1^{\uparrow}e^{-ik_1\left(x-x_1\right)}\right]\\
\fl 
\Psi_L^{\downarrow\left(\uparrow\right)}\left(-\infty<x<x_1\right)&&=\frac{8q_1q_2\sqrt{k_1}\left(k_3-k_4\right)\sin\theta}{den}e^{-ik_2\left(x-x_1\right)}\\
\fl 
\Psi_L^{\uparrow\left(\uparrow\right)}\left(x_1<x<x_2\right)&&=-\frac{2i}{den}\sqrt{\frac{k_1q_1}{q(x)}}\left(E\left(x_2,x\right)\left[\Psi\left(q_1,k_2,q_2,k_4\right)\left(q_2+ik_3\right)\left(1+\cos\theta\right)\right.\right.\\
&&\left.\left.+\Psi\left(q_1,k_2,q_2,k_3\right)\left(q_2+ik_4\right)\left(1-\cos\theta\right)\right]\right.\\
&&\left.+E^{-1}\left(x_2,x\right)\left[\Psi\left(q_1,k_2,q_2,k_4\right)\left(q_2-ik_3\right)\left(1+\cos\theta\right)\right.\right.\\
&&\left.\left.+\Psi\left(q_1,k_2,q_2,k_3\right)\left(q_2-ik_4\right)\left(1-\cos\theta\right)\right]\right)\\
\fl 
\Psi_L^{\downarrow\left(\uparrow\right)}\left(x_1<x<x_2\right)&&=\frac{4q_2}{den}\sqrt{\frac{k_1q_1}{q(x)}}\left(k_3-k_4\right)\sin\theta\left[E\left(x_1,x\right)\left(q_1-ik_2\right)+E^{-1}\left(x_1,x\right)\left(q_1+ik_2\right)\right]\\
\fl 
\Psi_L^{\uparrow\left(\uparrow\right)}\left(x_2<x<\infty\right)&&=-\frac{4i}{den}\sqrt{k_1q_1q_2}\left[e^{ik_3\left(x-x_2\right)}\Psi\left(q_1,k_2,q_2,k_4\right)\left(1+\cos\theta\right)\right.\\
&&\left.+e^{ik_4\left(x-x_2\right)}\Psi\left(q_1,k_2,q_2,k_3\right)\left(1-\cos\theta\right)\right]\\
\fl 
\Psi_L^{\downarrow\left(\uparrow\right)}\left(x_2<x<\infty\right)&&=-\frac{4i}{den}\sqrt{k_1q_1q_2}\left[e^{ik_3\left(x-x_2\right)}\Psi\left(q_1,k_2,q_2,k_4\right)-e^{ik_4\left(x-x_2\right)}\Psi\left(q_1,k_2,q_2,k_3\right)\right]\sin\theta
\end{eqnarray*}

Electrons initially in the right electrode have the following wavefunctions along the structure :

\begin{eqnarray*}
\fl \Psi_R^{\uparrow\left(\uparrow\right)}(-\infty<x<x_1)=-\frac{8i}{den}\sqrt{q_1q_2k_3}\Psi(q_1,k_2,q_2,k_4)\cos\frac{\theta}{2} e^{-ik_1(x-x_1)}\\
\fl \Psi_R^{\downarrow\left(\uparrow\right)}(-\infty<x<x_1)=-\frac{8i}{den}\sqrt{q_1q_2k_3}\Psi(q_1,k_1,q_2,k_4)\sin\frac{\theta}{2} e^{-ik_2(x-x_1)}\\
\fl \Psi_R^{\uparrow\left(\uparrow\right)}(x_1<x<x_2)=-\frac{4i}{den}\sqrt{\frac{k_3q_2}{q(x)}}\Psi(q_1,k_2,q_2,k_4)\cos\frac{\theta}{2}\left[E(x_1,x)(q_1-ik_1)+E^{-1}(x_1,x)(q_1+ik_1)\right]\\
\fl \Psi_R^{\downarrow\left(\uparrow\right)}(x_1<x<x_2)=-\frac{4i}{den}\sqrt{\frac{k_3q_2}{q(x)}}\Psi(q_1,k_1,q_2,k_4)\sin\frac{\theta}{2}\left[E(x_1,x)(q_1-ik_2)+E^{-1}(x_1,x)(q_1+ik_2)\right]\\
\fl \Psi_R^{\uparrow\left(\uparrow\right)}(x_2<x<\infty)=\cos\frac{\theta}{2}\frac{1}{\sqrt{k_3}}\left[e^{-ik_3(x-x_2)}-r_3^{\uparrow}e^{ik_3(x-x_2)}\right]+\sin\frac{\theta}{2}\frac{\sin\theta}{\sqrt{k_3}}\frac{8q_1q_2k_3(k_1-k_2)}{den}e^{ik_4(x-x_2)}\\
\fl \Psi_R^{\downarrow\left(\uparrow\right)}(x_2<x<\infty)=\sin\frac{\theta}{2}\frac{1}{\sqrt{k_3}}\left[e^{-ik_3(x-x_2)}-r_3^{\uparrow}e^{ik_3(x-x_2)}\right]-\cos\frac{\theta}{2}\frac{\sin\theta}{\sqrt{k_3}}\frac{8q_1q_2k_3(k_1-k_2)}{den}e^{ik_4(x-x_2)}
\end{eqnarray*}

To obtain $\Psi^{\downarrow\left(\downarrow\right)}$ and $\Psi^{\uparrow\left(\downarrow\right)}$ from $\Psi^{\uparrow\left(\uparrow\right)}$ and $\Psi^{\downarrow\left(\uparrow\right)}$, $\theta$ must be replaced by $-\theta$ and $k_1$ ($k_3$) by $k_2$ ($k_4$) in the above formulae.

\end{document}